\documentclass[preprint,showpacs,superscriptaddress]{revtex4}
\usepackage[dvips]{hyperref}
\usepackage{url}
\usepackage{epsfig}
\usepackage{pst-plot,url,epsf}

\newcommand{\beq}{\begin{equation}}
\newcommand{\eeq}{\end{equation}}
\newcommand{\bea}{\begin{eqnarray}}
\newcommand{\eea}{\end{eqnarray}}

\newcommand{\epm}{e^+e^-}

\newcommand{\ra}{\rightarrow}
\newcommand{\ga}{\gamma}

\newcommand{\eett}{e^+ e^- \ra t \bar{t}}
\newcommand{\ttbar}{t\bar{t}}

\newcommand{\sixf}{b f_1 \bar{f'_1} \bar{b} f_2 \bar{f'_2}}
\newcommand{\eesixf}{e^+ e^- \ra b f_1 \bar{f'_1} \bar{b} f_2 \bar{f'_2}}
\newcommand{\eebnmbdu}{e^+ e^- \ra b \nu_{\mu} \mu^+ \bar{b} d \bar{u}}

\newcommand{\AmS}{{\protect\the\textfont2
  A\kern-.1667em\lower.5ex\hbox{M}\kern-.125emS}}
\newcommand{\eebudbdu}{e^+ e^- \ra b u \bar{d} \bar{b} d \bar{u}}

\def \uo{\ensuremath { u(p_2) \,\, }}
\def \ut{\ensuremath { \bar{u}(p_3) \,\, }}
\def \vo{\ensuremath { v(p_4) \,\, }}
\def \vt{\ensuremath { \bar{v}(p_1) \,\, }}
\def\unity{{\rm 1\mskip-4.25mu l}}
\def \ps {p\hspace{-0.43em}/}
\newcommand{\fm}[2]{\bar{F}_{#1}^{#2}}
\newcommand{\fh}[2]{\hat{F}_{#1}^{#2}}
\newcommand{\fc}[2]{\bar{F}_{#1 B}^{#2^*}}
\def \dcs{\frac{\mathrm{d}\sigma}{\mathrm{d}\cos{\vartheta}}}
\def \eg{e.g.}

\begin{document}

\preprint{\tt DESY 03-178} \preprint{\tt SFB/CPP-03-53}

\title{Towards High Precision Predictions 
       for Top Quark Pair Production and Decay at a Linear Collider
\thanks{Presented by K. Ko\l odziej at the XXVII International Conference of 
        Theoretical Physics ``Matter to the Deepest: Recent Developments 
        in Physics of Fundamental Interactions'',
        Ustro\'n, Poland, September 15--21, 2003.}
\thanks{Work supported in part by the Polish State Committee for Scientific 
        Research (KBN) under contract No. 2~P03B~045~23 and by European 
        Commission's 5-th Framework contract  HPRN-CT-2000-00149.}%
\vspace{0.5cm}}

\author{Adam Biernacik}
\affiliation{Institute of Physics, University of Silesia, ul. Uniwersytecka 4, PL-40007 Katowice, Poland}%
\author{Karol Ko\l odziej}
\affiliation{Institute of Physics, University of Silesia, ul. Uniwersytecka 4, PL-40007 Katowice, Poland}%
\author{Alejandro Lorca}
\affiliation{Deutsches Elektronen-Synchrotron, DESY, Platanenallee 6, D-15738 Zeuthen, Germany}%
\author{Tord Riemann}
\affiliation{Deutsches Elektronen-Synchrotron, DESY, Platanenallee 6, D-15738 Zeuthen, Germany}

\date{October 2003\\\vspace{1cm}}

\begin{abstract}
We report on the progress in work on improving precision of the standard
model theoretical predictions for the top quark pair production and 
decay into six fermions at a linear collider. 
Two programs have been combined into a single Monte Carlo program:
{\tt eett6f}, a MC program
for $\epm \ra 6f$, and {\tt topfit}, a program for electroweak radiative 
corrections to $\epm \ra t{\bar t}$.
The MC program is described and preliminary numerical results are shown.
\end{abstract}

\pacs{12.15.-y, 13.40.Ks, 14.65.Ha}

\maketitle

\section{INTRODUCTION}
In order to disentangle possible new physics effects 
that may be revealed in the process of top quark pair production 
\bea
\label{eett}
         \epm \ra t \bar{t}  
\eea
at a linear collider 
from predictions of the standard model (SM), one needs to know the latter with 
high precision, possibly at the level of a few \mbox{per mill \cite{NLC}}.
That high precision level requires taking into account radiative corrections.
Because the top (and antitop) quark of reaction
(\ref{eett}) decays into 
a $b$ (and $\bar{b}$) and fermion-antifermion pairs,
we should study the 6 fermion reactions of the form
\bea
\label{ee6f}
     \eesixf, 
\eea
where $f_1, f'_2 =\nu_{e}, \nu_{\mu}, \nu_{\tau}, u, c$ and 
$f'_1, f_2 = e^-, \mu^-, \tau^-, d, s$.

Reactions (\ref{ee6f}) receive contributions from 
typically several
hundred Feynman diagrams already at the tree level, \eg
the hadronic reaction $\eebudbdu$ in the unitary gauge,
neglecting the Higgs boson couplings 
to ferm\-ions lighter than a $b$-quark, 
gets contributions from 1484 Feynman diagrams. 
This is the main reason why the calculation of the full $\cal{O}(\alpha)$ 
radiative corrections
to any of reactions (\ref{ee6f}) seems not to be feasible at the moment.

Therefore, in order to improve the precision of the lowest order predictions 
one should try to include at least the leading radiative effects, such as
initial state radiation (ISR) and factorizable radiative corrections
to the process of the on-shell top quark pair production (\ref{eett})
and to the decay of the top and antitop quarks.

In the present lecture, we report on the progress 
of the SM theoretical predictions for the top quark pair 
production and decay into six fermions at a linear collider,
paying special attention to the precision improvement. 

We will show a sample
of results on reactions (\ref{ee6f}) including some higher order
effects including the ISR correction obtained with
{\tt eett6f} \cite{eett6f}, a Monte Carlo program for reactions 
$\epm \ra$ 6 fermions 
relevant for the top production.
Further, we introduce the one--loop electroweak corrections
to reaction (\ref{eett}) obtained with {\tt topfit}
\cite{Fleischer:2003kk},  
a numerical package for calculating electroweak radiative corrections
to the on-shell pair production of a massive fermion--antifermion pair.
The two programs have been combined into a single Monte Carlo program.

\section{CALCULATING ISR WITH {\sc eett6f}}
The ISR correction has been implemented recently into 
{\tt eett6f} \cite{KK} in the leading logarithmic (LL)
approximation utilizing the structure function approach. In this
approach, the corrected
differential cross section ${\rm d}\, \sigma^{LL}(p_1,p_2)$ of
any reaction (\ref{ee6f}) reads
\bea
\label{LL}
  {\rm d}\, \sigma_{\rm Born + ISR}\left(p_1,p_2\right)=
\int_0^1 {\rm d}\, x_1 \int_0^1 {\rm d}\, x_2 \,
         \Gamma_{ee}^{LL}\left(x_1,Q^2\right)
\Gamma_{ee}^{LL}\left(x_2,Q^2\right) \nonumber \\ 
\times {\rm d}\, \sigma_{\rm Born}\left(x_1 p_1,x_2 p_2\right),
\eea
where $p_1$ ($p_2$) is the four momentum of a positron (electron),
$x_1$ ($x_2$) is the fraction of the initial momentum of the positron 
(electron) that remains after emission of a collinear photon and 
${\rm d}\, \sigma_{\rm Born}(x_1 p_1,x_2 p_2)$ is the lowest order cross 
section calculated at the reduced four momenta of the positron and electron. 
The structure function $\Gamma_{ee}^{LL}\left(x,Q^2\right)$ is given
by Eq.~(67) of \cite{Beenakker}, with {\tt `BETA'} choice for non-leading
terms. The splitting scale $Q^2$, which is not fixed in the LL approximation
is chosen to be equal $s=(p_1+p_2)^2$.

The ISR corrected cross sections, 
$\sigma_{\rm Born + ISR}$, of different channels 
of (\ref{ee6f}) obtained with the current version of {\tt eett6f} are 
compared with the results of {\sc Lusifer} \cite{Lusifer} in
Table~1 \cite{KK}, where also the corresponding lowest order total cross 
sections 
$\sigma_{\rm Born}$ are compared. The reader is referred to
\cite{didi} for a comparison of the lowest order predictions
with other existing MC programs.
The input parameters are the same as in \cite{Lusifer} and cuts are
given by
\bea
\label{cuts}
\begin{array}[b]{rlrlrlrl}
\theta (l,e^{\pm})> & 5^\circ, & \theta (q,e^{\pm})> & 5^\circ, &
\theta( l, l')> & 5^\circ, & \theta (l, q)> & 5^\circ,
\end{array}\nonumber \\
\begin{array}[b]{rlrlrl}
E_l > & 10\;{\rm GeV}, & E_q> & 10\;{\rm GeV}, &
 m(q,q') > & 10\;{\rm GeV}, \qquad
\end{array}  
\eea
where $\theta(i,j)$ is the angle between particles $i$ and $j$
in the laboratory system, $q$ and $l$ denote a quark and a final
state charged lepton, respectively, and $m(q,q')$ is the invariant mass
of a $qq'$ quark pair. 

\begin{table}
\label{tab1}
\caption{Comparison of the cross sections of different channels 
of (\ref{ee6f}) at $\sqrt{s}=500$~GeV of {\sc Lusifer} \cite{Lusifer} and 
{\tt eett6f} \cite{eett6f,KK}. 
The input parameters and cuts, see Eqs.~(\ref{cuts}) are the same as 
in \cite{Lusifer}. All cross sections are in fb.
The number in parenthesis show the uncertainty of the last decimals.}
\begin{center}
\begin{tabular}{|c|c|c|c|c|}
\hline
\rule{0mm}{7mm}  & \multicolumn{2}{c|}{\tt LUSIFER}
              & \multicolumn{2}{c|}{\tt eett6f}\\[2mm]
\cline{2-5}
\rule{0mm}{7mm} $\epm \ra$  & $\sigma_{\rm Born}$  & $\sigma_{\rm Born + ISR}$
        & $\sigma_{\rm Born}$ & $\sigma_{\rm Born + ISR}$  \\[2mm]
\hline
\rule{0mm}{7mm} $b\,\nu_e\,\epm \bar{\nu}_e \,\bar{b}$ 
       & 5.8530(68) & 5.6465(70) & 5.8622(63) & 5.6441(67) \\[1.5mm]
$b\,\nu_e \,e^+\mu^-\bar{\nu}_\mu\,\bar{b}$ & 5.8188(45) & 5.6042(38)
        & 5.8189(37) & 5.6075(59) \\[1.5mm]
$b\,\nu_{\mu}\,\mu^+\mu^-\bar{\nu}_{\mu}\,\bar{b}$ & 5.8091(49) & 5.5887(36)
        & 5.8065(33) & 5.5929(54) \\[1.5mm]
$b\,\nu_{\mu}\,\mu^+\tau^-\bar{\nu}_{\tau}\,\bar{b}$ & 5.7998(36) & 5.5840(40)
        & 5.7992(32) & 5.5844(33) \\[1.5mm]
$b\,\nu_{\mu}\,\mu^+ d\,\bar{u}\,\bar{b}$ & 17.171(24) & 16.561(24) 
      & 17.213(23) & 16.569(17) \\
without QCD: & 17.095(11) & 16.454(10) 
    & 17.106(15) & 16.457(16) \\[1.5mm]
$b\,\nu_{e}\,e^+ d\,\bar{u}\, \bar{b}$ & 17.276(45) & 16.577(21) 
      & 17.301(26) & 16.741(43)\\
without QCD: & 17.187(21) & 16.511(12) 
    & 17.149(16) & 16.522(17) \\[1.5mm]
\hline
\end{tabular}
\end{center}
\end{table}

\begin{figure}[ht]
\begin{center}
\setlength{\unitlength}{1mm}
\begin{picture}(35,35)(46,-50)
\rput(5.3,-6){\scalebox{0.45 0.45}{\epsfbox{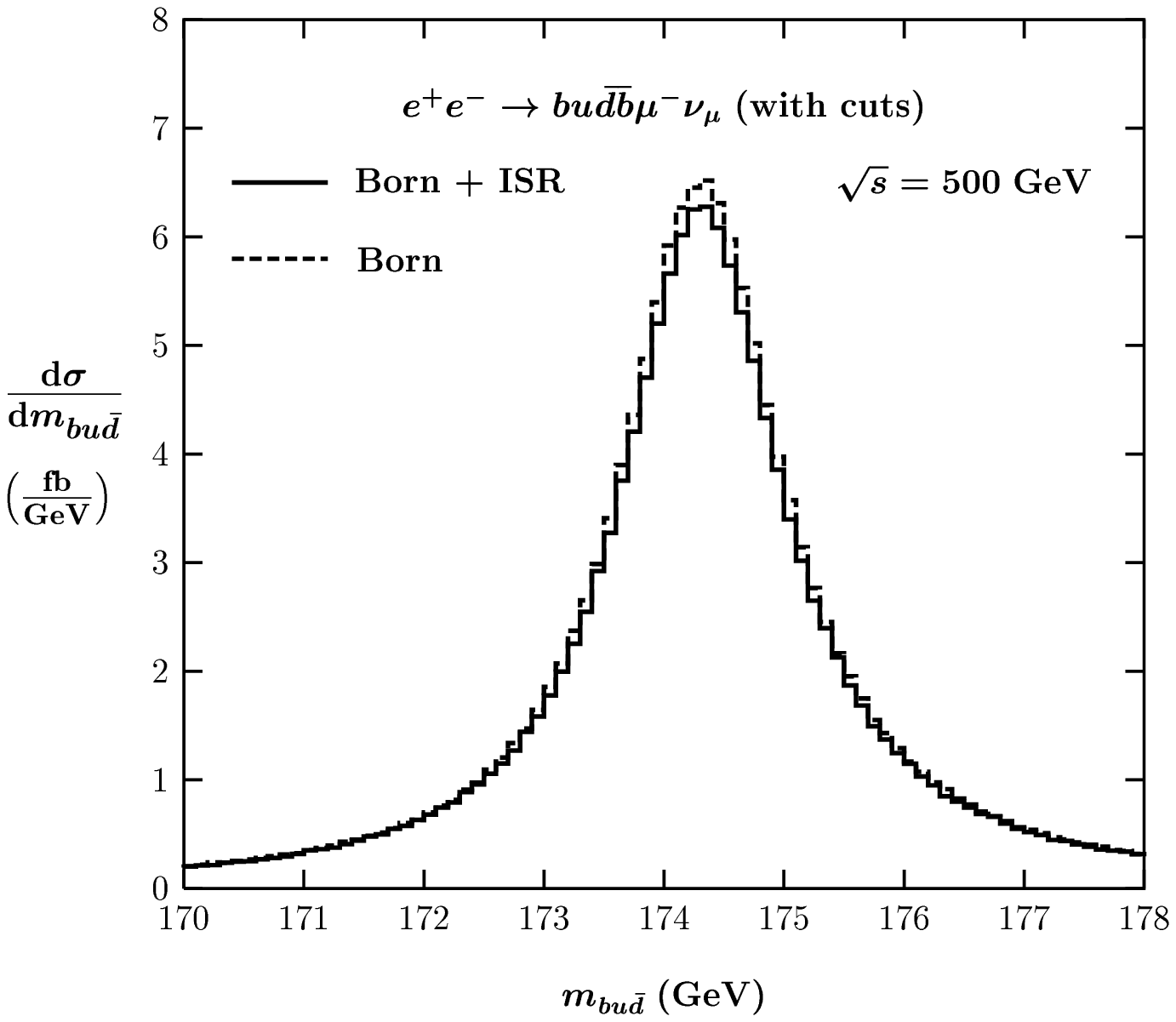}}}
\end{picture}
\begin{picture}(35,35)(18,-50)
\rput(5.3,-6){\scalebox{0.45 0.45}{\epsfbox{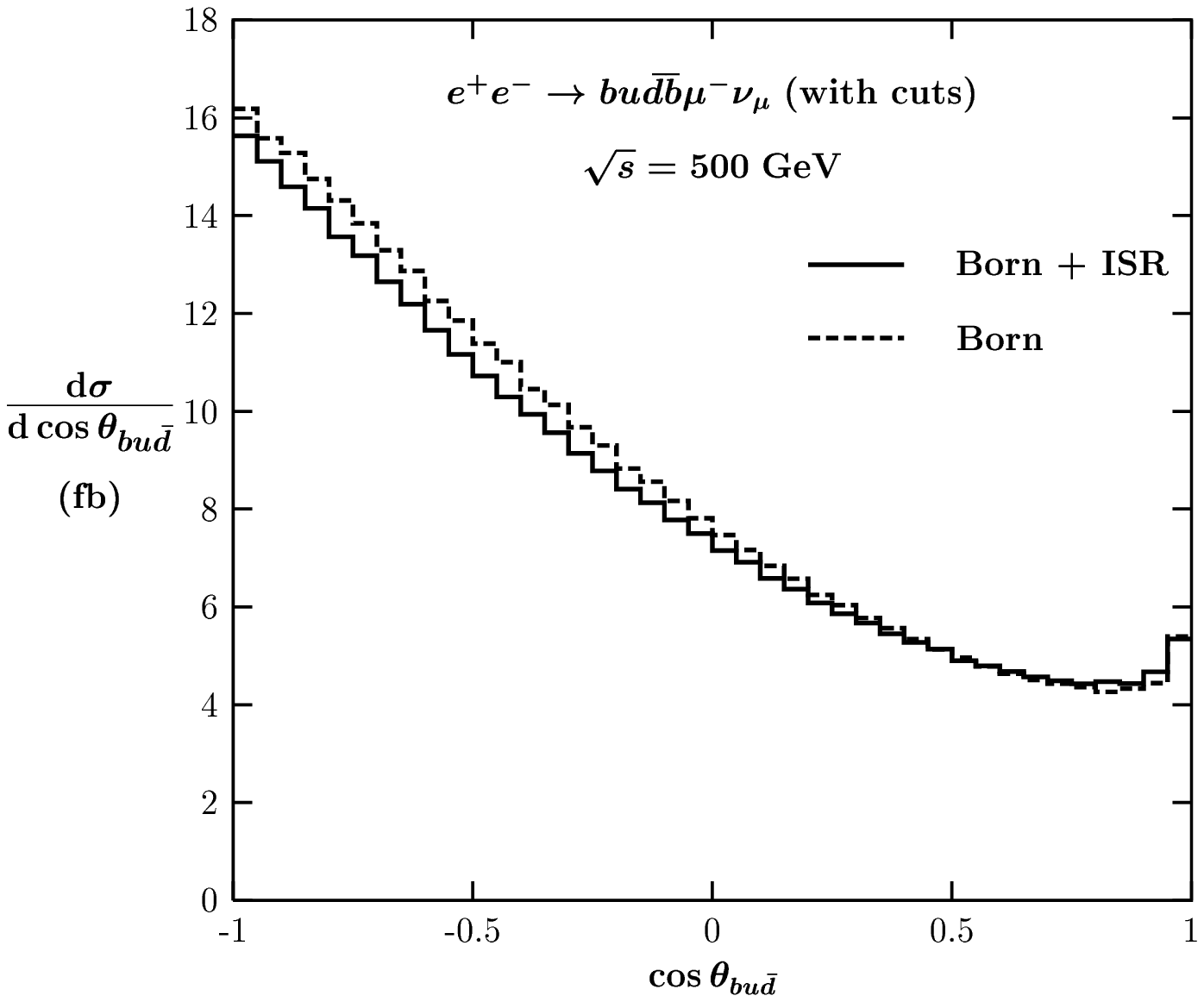}}}
\end{picture}
\end{center}
\vspace*{3.5cm}
\caption{Invariant mass (left) and angular (right) distributions 
         of a $\bar{b}d\bar{u}$-quark triple at $\sqrt{s}=500$ GeV
         with cuts.}
\label{fig0}
\end{figure}

The results of both programs
agree nicely, basically within one standard deviation, except for
$\epm \ra b\,\nu_{e}\,e^+ d\,\bar{u}\, \bar{b}$, where the deviation 
is bigger, amounting to about 3 standard deviations. The discrepancy 
should be clarified by fine tuned comparisons.
Differential cross sections obtained
with {\tt eett6f} that are plotted in Fig.\ref{fig0} agree with those presented 
in \cite{Lusifer} within
accuracy of the plots.

\section{CALCULATING ONE-LOOP ELECTROWEAK CORRECTIONS WITH {\sc topfit}}
{\tt topfit} \cite{Fleischer:2003kk} is a numerical package for
calculating the complete 
one--loop corrections to top-pair production (\ref{eett})
at a linear $\epm$ collider
in the continuum energy region. 
The corrections are represented in terms of
six independent form factors, {which} is suitable for implementation
in the Monte Carlo generator.
We recall the normalization of the differential cross section
\bea
\label{normal}
\frac{{\rm d}\,\sigma(\eett)}{{\rm d}\,\cos\vartheta}
= \frac{\beta_t} {32\pi \, s} \,  
\sum_{\mathrm{conf}}\left[ | {\cal M}_B |^2 
+2 {\rm \Re e}\left({\cal M}^*_B \delta {\cal M} \right)
\right]\,
\eea
where $\vartheta$ is the antitop production angle with respect to the
initial $e^+$ beam, $\beta_t = \left(1-{4\,m_t^2}/{s}\right)^{1/2}$
and the sum stands for the spin and color configuration of external fermions. 
There are four nonvanishing form factors ${\rm F}_{1B}^{ab}$ in Born 
approximation
\bea
\label{Bornamp}
 \, {\cal M}_{B} = 
\sum_{a,b=\{1,5\}} {\rm F}_{1B}^{\,ab} \,\,    {\cal M}_{1,{ab}},
\eea
while the one--loop correction $\delta {\cal M}$ to the amplitude
can be expressed in terms of 
six independent form factors: $\fh{1}{ab}, \; a,b=\{1,5\}$, plus
additionally $\fh{3}{11}$ and $\fh{3}{51}$:
\bea
\label{damp}
 \delta {\cal M} = 
\sum_{a,b=\{1,5\}} \fh{1}{ab} \,\,    {\cal M}_{1,{ab}} +  
 \fh{3}{11} \,\,    {\cal M}_{3,{11}} +  
 \fh{3}{51} \,\,    {\cal M}_{3,{51}}.
\eea

The invariant amplitudes in Eqs.~(\ref{Bornamp}) and (\ref{damp}) are
given by
\bea
\label{amps}
i \, {\cal M}_{1,\,{ab}} & =& [\vt \ga^{\mu}\, g_a \, \uo] \,\, [\ut
\ga_{\mu}\, g_b \, \vo] , \quad  a,b=\{1,5\},\nonumber\\
i \, {\cal M}_{3,{11}} & = & - [\vt  \ps_3\, \unity \, \uo] \,\, [\ut \unity \, \vo]  ,\\
i \, {\cal M}_{3,{51}} & = &-[\vt \ps_3\, \ga_5 \, \uo] \,\, 
[\ut \unity \, \vo].\nonumber
\eea

The normalization of the form factors {is} chosen such that 
the pure photonic Born
amplitude becomes \mbox{$F_{1 B}^{11,{\gamma}} = e^2Q_eQ_t/s$};
(see \cite{Fleischer:2003kk} for the details on the analytic expressions).

%

The resulting corrected differential cross section formula can then be written in terms
of the computed form factors, the top quark color factor ($c_t$) and
the kinematical Mandelstam variables $s$,$t$ and $u$ ($s+t+u=2m_t^2$):
\bea
\label{explicitdcs}
\frac{{\rm d}\,\sigma}{{\rm d}\,\cos\vartheta}&=&
\frac{\beta_t c_t}{s}  \frac{\alpha^2 \pi}{s^2}  {\rm \Re e} \bigg\{
4 s m_t^2 \left( \fm{1}{11} \fc{1}{11} - \fm{1}{15} \fc{1}{15} +
  \fm{1}{51} \fc{1}{51} - \fm{1}{55} \fc{1}{55}\right) 
\nonumber\\
&&\hspace{-33pt} + 2(u^2-t^2-2m_t^2(u-t)) \left( \fm{1}{11} \fc{1}{55} + \fm{1}{55} \fc{1}{11} +
  \fm{1}{51} \fc{1}{15} + \fm{1}{15} \fc{1}{51}\right) 
\nonumber \\
&&\hspace{-33pt} + 2(u^2+t^2+2m_t^2(s-m_t^2))
\sum_{a,b=\{1,5\}}\fm{1}{ab}\fc{1}{ab} 
\nonumber  \\
&&\hspace{-33pt} + 4(ut-m_t^4)\left( \fm{3}{11} \fc{1}{11} + \fm{3}{51} \fc{1}{51}
\right) \bigg\},
\eea
where the dimensionless {\em barred} form factors are defined for
convenience as 
\bea
\label{ffbardef}
\fc{1}{ab} &\equiv& \frac{s}{e^2} F_{1 B}^{ab^*}, \nonumber\\
\fm{1}{ab} &\equiv& \frac{s}{e^2} \left (\frac{1}{2} F_{1 B}^{ab} +
  \fh{1}{ab} \right), \\
\fm{3}{ab} &\equiv& \frac{s}{e^2} m_t \fh{3}{ab}. \nonumber
\eea

{\tt topfit} is rather flexible and includes also real photonic
bremsstrahlung. In order to fit the needs of the MC program described here, some specific
values of flags have been chosen:
\mbox{{\tt IWEAK} = 1} (inclusion of pure weak corrections), 
\mbox{{\tt IQED} = {\tt IQEDAA} = 0}. The latter two settings switch off the
calculation of photonic corrections due to bremsstrahlung ({\tt IQED}, with the related
virtual corrections) and of the fermionic vacuum polarization effects ({\tt IQEDAA}). 
 The sample corrections were computed using the same input parameters as in
\cite{Hahn:2003ab}.
In Table \ref{topfitaa}, form factors are given for  $\sqrt{s}=500$ GeV
and $\cos{\vartheta}=0.7$, and in Table \ref{topfitbb} some differential
cross section values for  $\sqrt{s}=500$ GeV.

\begin{table}[tbp]
\caption{Born and weak one--loop form
  factors for $\epm \ra t{\bar t}$ at $\sqrt{s}=500$ GeV and
  $\cos{\vartheta}=0.7$.  }
\label{topfitaa}
{\small
$$
\begin{array}{|r|r@{\cdot}lc|r@{\cdot}lr@{\cdot}l|}
\hline
& \multicolumn{3}{c|}{\textrm{Born F.F.} \quad {\rm (GeV^{-2})}} & \multicolumn{4}{c|}{\textrm{Weak
  one--loop F.F.}\quad {\rm (GeV^{-2})}}\\
               & \multicolumn{2}{c}{\rm \Re e} & {\rm \Im m} &
               \multicolumn{2}{c}{\rm \Re e} & \multicolumn{2}{c|}{\rm \Im m} \\
\hline
\hat{F}_{1}^{11} &-2.50926472&10^{-7}&0& 1.08874859&10^{-8} &-2.55015622&10^{-9}\\
\hat{F}_{1}^{15} & 1.56200966&10^{-8}&0&-9.75979648&10^{-9} &-9.02716758&10^{-9}\\
\hat{F}_{1}^{51} & 5.62400131&10^{-8}&0&-7.00614109&10^{-9} &-6.39336661&10^{-9}\\
\hat{F}_{1}^{55} &-1.37479846&10^{-7}&0&-1.11566191&10^{-9} & 7.86314816&10^{-9}\\
m_t\hat{F}_{3}^{11} &\multicolumn{2}{c}{0}&0& 8.14431322&10^{-10}&-8.44590997&10^{-10}\\
m_t\hat{F}_{3}^{51} &\multicolumn{2}{c}{0}&0&-9.09997715&10^{-10}& 4.88604910&10^{-10}\\
\hline
\end{array}
$$
}
\end{table}

\begin{table}
\caption{
Differential cross sections for $\epm \ra t{\bar t}$ at
  $\sqrt{s}=500$GeV.}
\label{topfitbb}
$$
\begin{array}{|r|c|c|}
\hline
 \cos\vartheta \phantom{\bigg|} &
 \dcs\Big|_{\textrm{Born}} \quad
   {\rm (pb)}&\dcs\Big|_{\textrm{Born+Weak}} \quad {\rm (pb)}\\
\hline
-0.9 & 0.1088391941 & 0.1011751626 \\
-0.7 & 0.1218770828 & 0.1136440560 \\
-0.5 & 0.1422750694 & 0.1332668988 \\
 0.0 & 0.2254704640 & 0.2115640646 \\
 0.5 & 0.3546664703 & 0.3300633143 \\
 0.7 & 0.4192250441 & 0.3883965473 \\
 0.9 & 0.4911437158 & 0.4528185418 \\
\hline
\end{array}
$$
\end{table}

\section{NUMERICAL RESULTS}
In the present section, we will show the results of implementation of
the one--loop electroweak form factors for on-shell top quark pair production
obtained with {\tt topfit} in the MC program {\tt eett6f}. 
Amplitudes (\ref{amps}) have been implemented in {\tt eett6f} with
the helicity amplitude method of \cite{KZ,JK}.
The actual numerical values of the form factors
$F_{1 B}^{ab}, \fh{1}{ab}, \; a,b=\{1,5\}$,
$\fh{3}{11}$ and $\fh{3}{51}$
are computed by {\tt topfit} and then transferred to {\tt eett6f}. 
Unfortunately, the direct calculation of the
one--loop corrections slows down the MC computation by more than a factor 1000.
In practice, this problem has been solved 
applying the following strategy: the
values of the one--loop form factors are computed for a given
c.m.s.\ energy for a grid of several hundred, or a few thousand different
values of cosine of the antitop scattering angle, $\cos\vartheta$. These
fixed values of the form factors are stored in the 
computer memory and then used for computation of the form factors at
any intermediate value of $\cos\vartheta$ by means of a linear interpolation 
between the neighbouring fixed values. The interpolation routine
has been tested and shown to be very precise and efficient.

As a first step we calculate the differential cross section of reaction 
(\ref{ee6f})
in the narrow width approximation for the top and antitop, which is
obtained by multiplying the one--loop corrected cross section (\ref{normal})
with the corresponding branching fractions leading to a specific final state 
of (\ref{ee6f})
\bea
\label{cstt}
{\rm d}\, \sigma\left(\epm \ra \ttbar \ra \sixf\right)= 
  {\rm d}\,\sigma(\eett)\qquad \qquad \nonumber\\
\times \; {\rm d}\,\Gamma\left(t \ra b f_1 \bar{f'_1}\right)\;
{\rm d}\, \Gamma\left(\bar{t} \ra \bar{b} f_2 \bar{f'_2}\right)/{\Gamma_t}^2.
\eea

This approach is gauge invariant and can be used as the reference for
any other way of implementation of the one--loop electroweak corrections
in the MC generator for a simulation of reactions (\ref{ee6f}) to
$\cal{O}(\alpha)$.

\begin{figure}[ht]
\begin{center}
\setlength{\unitlength}{1mm}
\begin{picture}(35,35)(45,-50)
\rput(5.3,-6){\scalebox{0.5 0.5}{\epsfbox{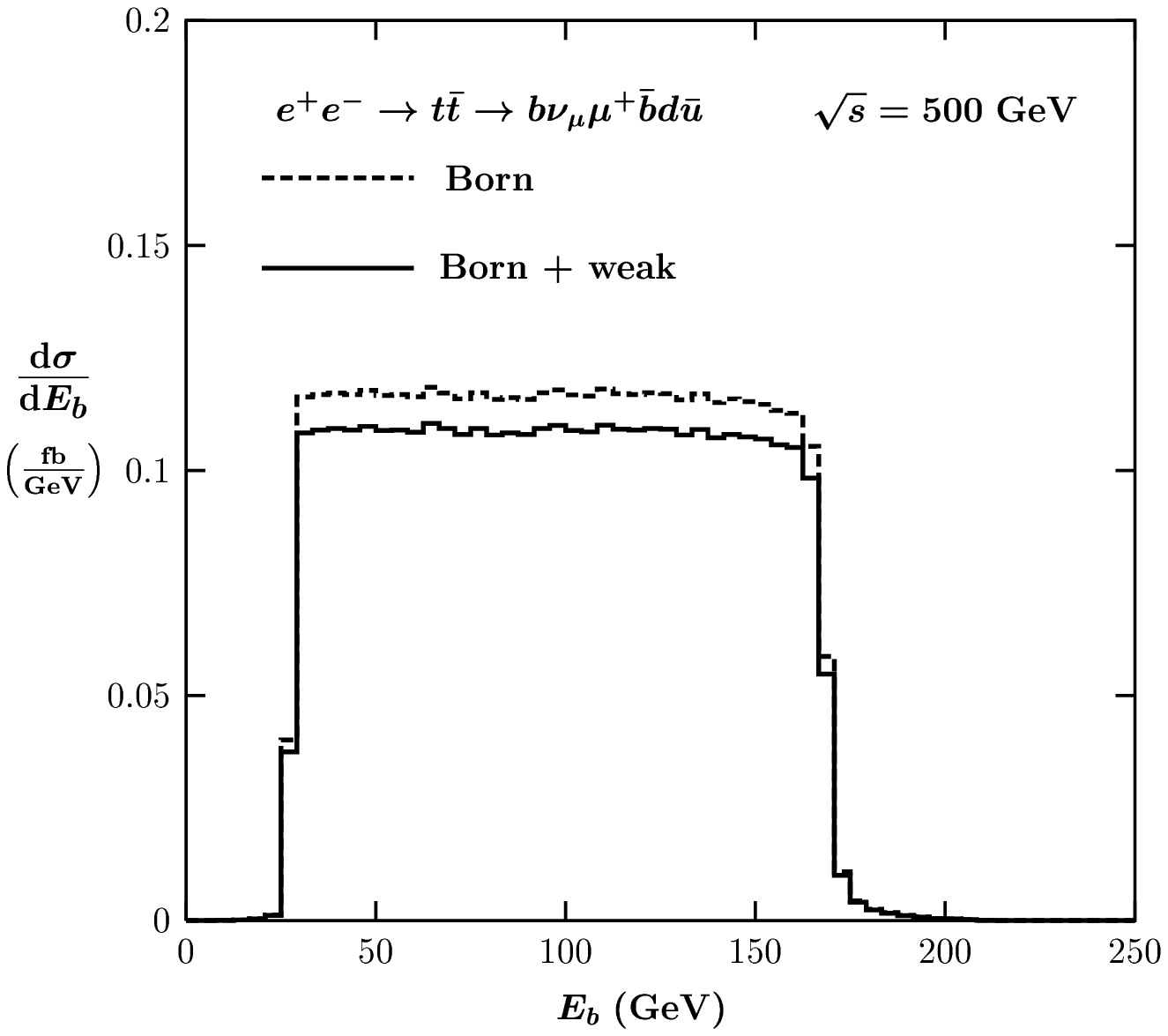}}}
\end{picture}
\begin{picture}(35,35)(20,-50)
\rput(5.3,-6){\scalebox{0.5 0.5}{\epsfbox{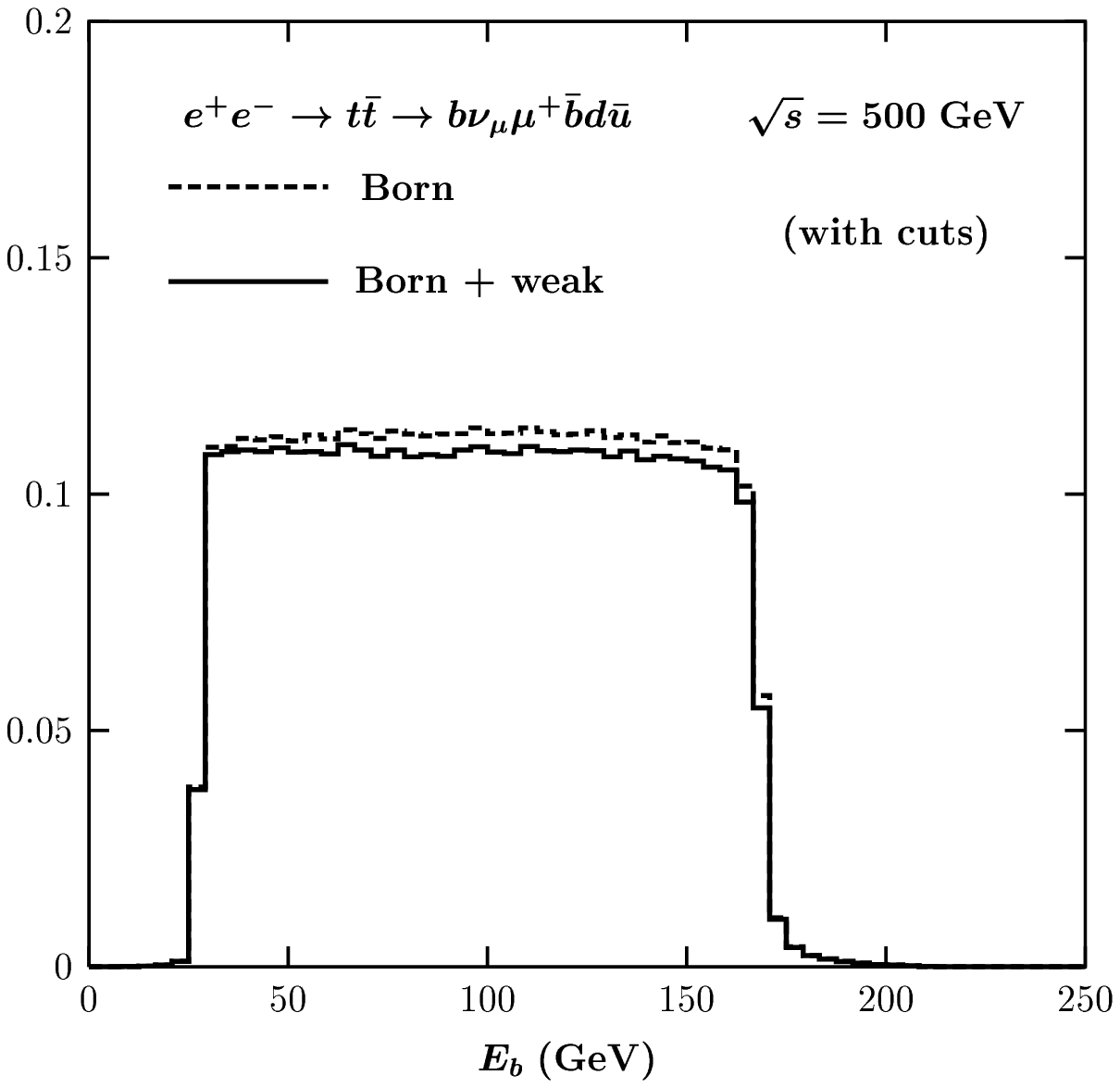}}}
\end{picture}
\end{center}
\vspace*{3.5cm}
\caption{Energy distributions of a $b$-quark at $\sqrt{s}=500$ GeV without 
cuts and with cuts.}
\label{fig3}
\end{figure}

\begin{figure}[ht]
\begin{center}
\setlength{\unitlength}{1mm}
\begin{picture}(35,35)(45,-50)
\rput(5.3,-6){\scalebox{0.5 0.5}{\epsfbox{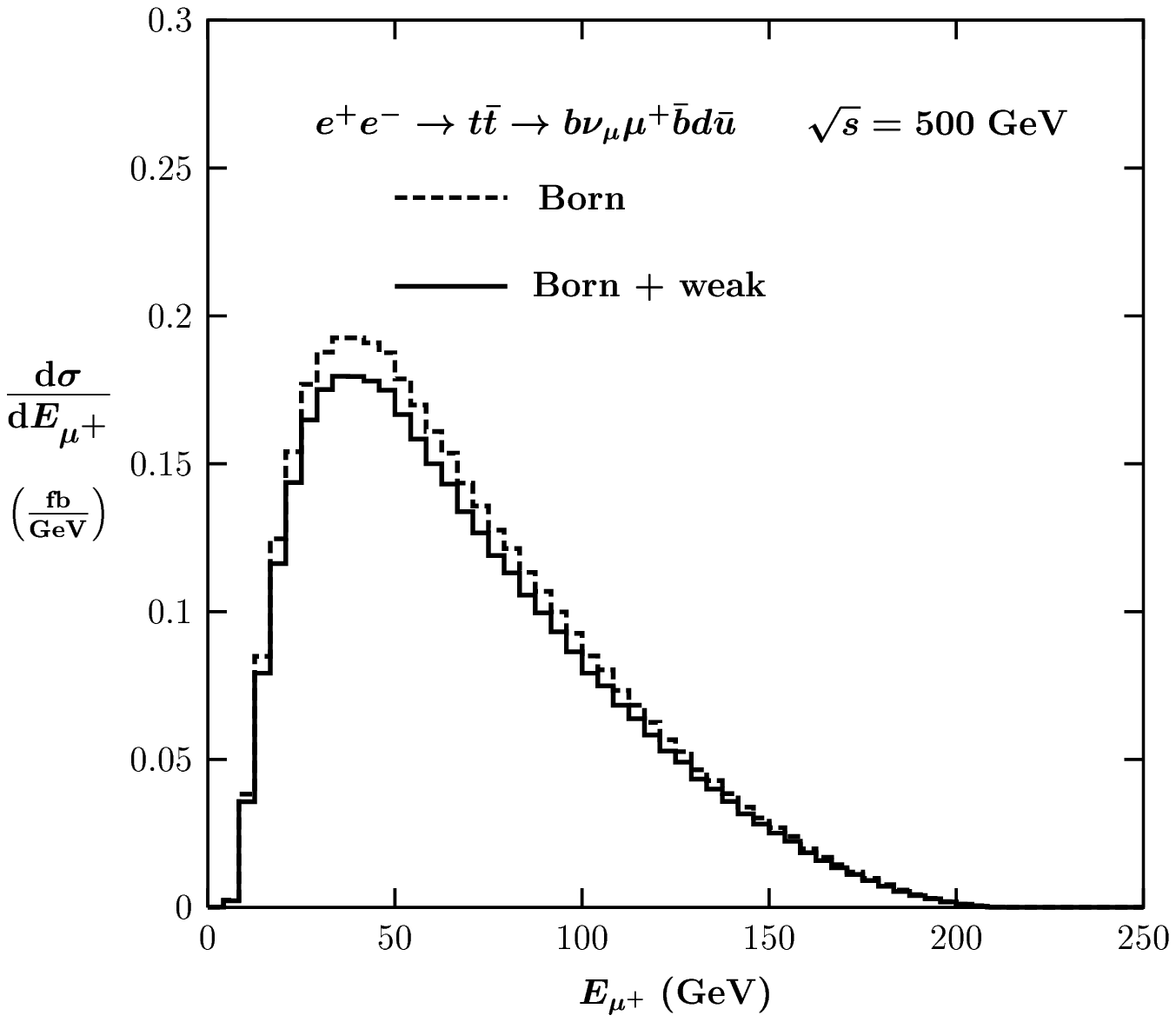}}}
\end{picture}
\begin{picture}(35,35)(20,-50)
\rput(5.3,-6){\scalebox{0.5 0.5}{\epsfbox{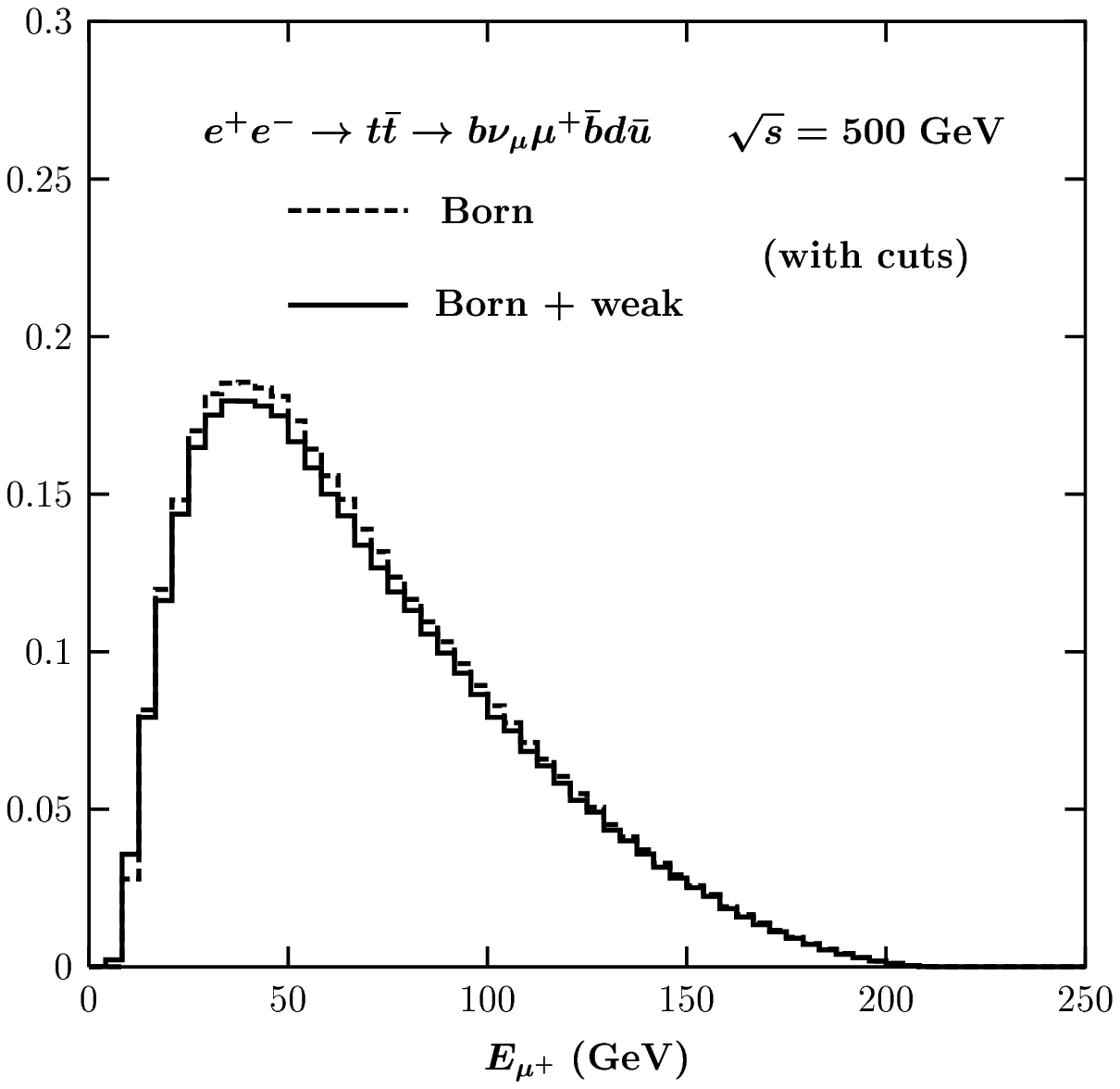}}}
\end{picture}
\end{center}
\vspace*{3.5cm}
\caption{Energy distributions of a $\mu^+$ at $\sqrt{s}=500$ GeV without 
cuts and with cuts.}
\label{fig4}
\end{figure}

Typical results for one specific channel of reaction (\ref{ee6f})
in the narrow top width approximation of 
Eq.~(\ref{cstt}) are shown in Figs.~\ref{fig3} and \ref{fig4}. The differential
cross sections of reaction $\eebnmbdu$ at $\sqrt{s}=500$~GeV without cuts 
are plotted on the
left hand side of Fig.~\ref{fig3}. Both the lowest order (dotted line) and one--loop 
corrected (solid line) differential cross sections in the $b$-quark energy 
are shown. The same differential cross sections with cuts of
conditions (\ref{cuts}) are plotted on the right hand side of Fig.~\ref{fig3}.
The corresponding differential cross sections 
in the energy of $\mu^+$ are plotted in Fig.~\ref{fig4}. 

We see that the quite substantial effect of the one--loop electroweak
correction on differential cross sections of
$\eebnmbdu$ at $\sqrt{s}=500$~GeV is reduced by the cuts (\ref{cuts}).
This kind of reduction is of course not desirable if one wants to
test the consistency of the SM at the quantum level. However, it might
be helpful in performing relatively fast numerical simulations of 
different channels of reaction (\ref{ee6f}) with the use of the MC generators
working only with the lowest order amplitudes.

\section{SUMMARY AND OUTLOOK}

{We have reported on the progress in our work on improving precision of SM
theoretical predictions for the top quark pair production and 
decay into six fermions at a linear collider. 
Apart from including the ISR effects in {\tt eett6f} \cite{KK},
we have created a MC program that combines
two programs {\tt eett6f} and {\tt topfit}. The program allows
for including one--loop electroweak corrections to the top quark pair
production 
in reactions (\ref{ee6f}) in the narrow width approximation for the top 
and antitop.}
The description of these corrections to the on-shell top pair
production has to be complemented by the corresponding ones to the
  decays of the top quarks \cite{Denner:1990ns}, in order to complete the
  double pole approximation. This is in preparation \cite{karoletal}.

\end{document}